\documentstyle[11pt]{article}

\newcommand{\bb}{\begin{equation}}
\newcommand{\ee}{\end{equation}}
\newcommand{\bqn}{\begin{eqnarray}}
\newcommand{\eqn}{\end{eqnarray}}

\begin{document}
\bibliographystyle{BKstyle}

\begin{titlepage}

\begin{flushright}

ULB--TH--00/08\\ DAMTP-2000-35 \\hep-th/0003170\\

\end{flushright}
\vfill

\begin{center} {\Large{\bf Couplings of gravity to
antisymmetric gauge fields}}
\end{center}
\vfill

\begin{center} {\large Xavier Bekaert$^{a}$, Bernard
Knaepen$^{a,b*}$ \\ and Christiane Schomblond$^a$}
\end{center}
\vfill

\begin{center}{\sl
$^a$ Facult\'e des Sciences, Universit\'e Libre de Bruxelles,\\
Campus Plaine C.P. 231, B--1050 Bruxelles, Belgium\\[1.5ex]

$^b$ DAMTP, Centre for Mathematical Sciences,\\ Wilberforce
Road, CAMBRIDGE CB3 0WA, UK

}\end{center}
\vfill

\begin{abstract}
We classify all the first-order vertices of gravity
consistently coupled to a system of
$2$-form gauge fields by computing the local BRST cohomology
$H(s\vert d)$ in ghost number $0$ and form degree $n$.  The
consistent deformations are at most linear in the
undifferentiated two-form, confirming the previous results of
\cite{Damour:1992ru} that geometrical theories constructed from
a nonsymmetric gravity theory  are physically inconsistent or
trivial. No assumption is made here on the degree of
homogeneity in the derivatives nor on the form of the gravity
action.
\end{abstract}

\vspace{5em}

\hrule width 5.cm
\vspace*{.5em}

{\small \noindent (*)Boursier post-doc 
de la fondation Wiener-Anspach (Belgique).}

\end{titlepage}

\section{Introduction}
\setcounter{equation}{0}
\setcounter{theorem}{0}

Long ago, the question has been addressed of constructing
geometric field theories of gravity from the non-symmetric
metric 
\begin{equation} G_{\mu\nu} \equiv g_{\mu\nu} + B_{\mu\nu} \; ,
\label{Ab}
\end{equation} where $g_{\mu\nu}$ is the symmetric part of
$G_{\mu\nu}$ and $B_{\mu\nu}$ is an antisymmetric tensor
field.  Such a model was first discussed in 1925 by Einstein
\cite{Einstein25} in an attempt to unify gravity and
electromagnetism,  and developed by himself and others
\cite{EinsteinStraus}. In some sense, modern
string theory has revived this attempt at geometric unification
of forces \cite{Scherk:1974mc}

But the couplings of a gauge field like a massless 2-form
$B_{\mu\nu}$ and a (symmetric) metric $G_{\mu\nu}$ are severely
restricted by consistency requirements,  thereby disqualifying
the old theory of Einstein and its modern versions (like the
``nonsymmetric gravity theory'' of Moffat
\cite{Moffat:1979tr}). The first proof of the physical
inconsistency of these models has been given by T. Damour, S.
Deser and J. McCarthy \cite{Damour:1992ru}. They
have shown that all ``geometric'' actions (see
\cite{Damour:1992ru} for a precise explanation of the sense of
this term) homogeneous of order two in the number of
derivatives violate standard physical requirements. In this
paper, we confirm their results while relaxing some of the
assumptions made by these authors.

To analyze the generic geometric models, they had to expand the
action in powers of $B_{\mu\nu}$ about a classical symmetric
background, and consider the merits of the resulting theory,
that means study the consistency of the deformations of a
theory of gravity coupled to a differential 2-form in terms of
standard physical criteria: absence of negative-energy
excitations and coherence of the degree-of-freedom content. The
first point of their argument is that the absence of
negative-energy excitations requires (whatever the
gravitational background) the $B$ expansion to begin with the
(quadratic) kinetic term $H_{\mu\nu\lambda}H^{\mu\nu\lambda}$
where $H_{\lambda\mu\nu} \equiv \partial_{\lambda}B_{\mu\nu} +
\partial_\mu B_{\nu\lambda} + \partial_\nu B_{\lambda\mu}$ is
the field strength. This action has the usual gauge invariance 
$\delta B_{\mu\nu} = \partial_{\mu} \epsilon_{\nu} -
\partial_{\nu}\epsilon_{\mu}$. Secondly, geometric actions
homogeneous in two derivatives generate powers of the
undifferentiated $B_{\mu\nu}$.  These higher-power terms
generically violate the gauge invariance of the leading kinetic
term.  Furthermore, it is even impossible to overcome this
problem by a deformation of the abelian gauge
invariance \cite{Damour:1992ru}.

A modern useful tool to study consistent deformations is the
BRST field-antifield formalism. It is known that if $s$
is the BRST operator of the free theory,  the first order
vertices of consistent deformations are constrained to belong
to the cohomological group
$H^n_0(s\vert d)$ of $s$ modulo $d$ in ghost number $0$ and
form degree $n$ (the dimension of spacetime)
\cite{Barnich:1993vg}. By using the powerful tool of
homological algebra, we completely determine all non trivial
consistent deformations at first order in the coupling constant
for a system of abelian 2-forms coupled to gravity for $n\geq
4$.\footnote{For $n<4$, the 2-forms do not carry any degree of
freedom, hence this case is not considered.}

Our result is that all non trivial first-order consistent
deformations are of the four following types:
\begin{itemize}
\item Strictly invariant polynomials in the curvatures
$R_{\mu\nu\rho\sigma}$ and $H^A_{\mu\nu\rho}$.
\item Couplings invariant under the abelian transformation up
to a divergence. They generalize the purely gravitational
deformations previously known in the literature (denoted as
${\cal{A}}_{chiral}$ in \cite{Brandt:1990et}) and the
Chern-Simons type self-coupling of the 2-forms (which exist
only for specific values of $n$).
\item Exotic couplings of the 2-form gauge fields to
gravitational Chern-Simons forms (see \cite{Brandt:1999xk}). 
These interactions deform the gauge transformations of the
free theory in a non trivial way.
\item The Freedman-Townsend coupling \cite{Freedman:1981us}
(only allowed for $n=4$), where the abelian gauge
transformations for the 2-forms are non trivially deformed.
\end{itemize} 

We notice that non trivial deformations of ``gravity + 2-form''
are at most linear in the undifferentiated $B_{\mu\nu}$ which
confirms the results of Damour et al. that geometric actions
including higher powers of the undifferentiated
$B_{\mu\nu}$ are physically inconsistent or trivial. No
assumption is made here on the degree of homogeneity in the
derivatives nor on the form of the gravity action (except that
it must define a normal theory, according to the
terminology of \cite{Barnich:1995db}).

In the next section, we determine all the cohomological class
of $H(s\vert d)$ in ghost number $0$ and form degree $n$ and we
relate those classes to the first-order consistent deformations
of the theory.

\section{The model}
\setcounter{equation}{0}
\setcounter{theorem}{0} 

Throughout the paper we shall work in
the vielbein formulation of gravity. Denoting the vielbein
fields by $e^{\ a}_\mu$, we define the inverse 
vielbeins, the metric, the Christoffel connection and the spin
connection through,
\begin{eqnarray} E_a^{\ \mu}e_\mu^{\ b}=\delta^a_b,\quad
e_\mu^{\ a}E_{a}^{\
\nu}=\delta^\nu_\mu,\quad g_{\mu\nu}=e_{\mu a}e_{\nu}^{\ a},
\quad g^{\mu\nu}=E^{a\mu}E_a^{\ \nu},
\end{eqnarray}
\begin{equation}
\partial_\mu e_\nu^{\ a}-\omega_{\mu b}^{\ \ a}e_\nu^{\ b} -
\Gamma_{\mu\nu}^{\ \ \rho}e_\rho^{\ a}=0,\quad 
\omega_{\mu }^{\ ab}=-\omega_{\mu }^{\ ba},\quad
\Gamma_{\mu\nu}^{\ \ \rho} =\Gamma_{\nu\mu}^{\ \ \rho},
\end{equation} where  Lorentz indices ($a,b,\ldots$) are raised
and lowered with the Minkowski metric.

As we pointed out in the introduction, our calculations are
valid for a wide range of gravity action. For simplicity, the
model we consider here is the standard Einstein-2-form system.
Its lagrangian reads,
\begin{equation} {\cal L}_0/e=\frac12 R
-\frac{1}{12}g^{\mu\lambda} g^{\nu\sigma} g^{\rho\theta}
H^A_{\mu\nu\rho}H_{A
\lambda\sigma\theta},\label{lagrangian}
\end{equation} where $e$ is the determinant of the vielbeins,
$R=R_{\mu\nu}^{\
\ ab}E_{a}^{\ \mu} E_{b}^{\ \nu}$ is the Riemann curvature and
$H^A_{\mu\nu\rho}$ are the $2$-form field strengths.

Because of the gauge invariance of the
classical action under the diffeomorphisms, local Lorentz
transformations and gauge transformations of the $2$-form
potentials ($B^A_{\mu\nu}\rightarrow B^A_{\mu\nu} +\partial_\mu
\epsilon^A_\nu -\partial_\nu \epsilon^A_\mu$), we are led to
introduce within the field-antifield formalism the following
set of ghosts: $\xi^\mu$ (diffeomorphisms ghosts), $C^{ab}$
(Lorentz ghosts), $B^A_{\mu}, B^A$ ($2$-form ghosts). 
The extra ghosts $B^A$ (``ghosts of ghosts") required
in the 2-form sector arise because the gauge transformations of
the $2$-form potentials are reducible. Indeed, they vanish when
the gauge parameters are set to
$\epsilon^A_\mu=\partial_\mu
\lambda^A$ where $\lambda^A$ are arbitrary functions.
In the sequel we will denote by $\Phi^\Gamma$ all the fields
and ghosts:
$
\{\Phi^\Gamma \}=\{ e^{\ a}_\mu, 
B^A_{\mu\nu},\xi^\mu,  C^{ab}, B^A_\mu, B^A\}$. The
corresponding antifields will be denoted by
$\Phi^*_\Gamma$:
$
\{\Phi^*_\Gamma \}=\{ e_{a}^{*\mu}, 
B^{*\mu\nu}_A,\xi^*_\mu, C^*_{ab}, B^{*\mu}_A, B^{*}_A\}$.
In terms of these variables the minimal solution of the BRST
master equation reads:
\begin{eqnarray} S&=&\int d^n x({\cal L}_0 -\xi^\mu
\Phi^*_\Gamma \partial_\mu
\Phi^\Gamma + e^{*\mu}_a (e_\nu^{\ a}\partial_\mu \xi^\nu +
e_\mu^{
\ b}C_b^{\ a}) \nonumber +
\frac12 C^*_K C^I C^J f_{IJ}^{\ \ K} \\ && \quad \quad \quad +
B^{*\mu\nu}_A (\partial_\mu B^A_\nu + \partial_\mu \xi^\rho
B^A_{\rho\nu}) - B^{*\mu}_A (\partial_\mu B^A + \partial_\mu
\xi^\rho B^A_\rho)).
\end{eqnarray} In the above expression the 
$f_{IJ}^{\ \ K}$ denote   the structure constants
of ${\cal G}_L$. We have also used
the notation,
$
C^I = C^{ab}.
$
The action of the BRST differential on the variables is then
defined as:
$ s\Phi^A = -\frac{\delta^R
S}{\delta \Phi^*_A}, s\Phi^*_A = \frac{\delta^R
S}{\delta \Phi^A}$.

In order to obtain the possible deformations of
(\ref{lagrangian}) we need to compute the local
BRST cohomology 
$H(s|d)$ in the algebra of local forms. These are by definition
linear combinations of spacetime forms $\omega^\tau (x,dx)$
with coefficients that are  local functions. By  local
functions it is meant functions which depend polynomially on 
the variables $\partial_{\mu_1 \ldots \mu_j} \Phi^\Gamma$ and 
$\partial_{\mu_1 \ldots \mu_j} \Phi^*_\Gamma$ except the
undifferentiated vielbeins $e_\mu^{\ a}$ for which a smooth and
regular dependence in an open neighborhood of some regular
background configuration ($\mbox{det}(e_\mu^{\ a})\not = 0$)
is allowed. We shall denote the algebra of local forms by
${\cal E}$ and the algebra of local functions by ${\cal A}$.
Thus:
${\cal E}= \Omega (M) \otimes {\cal A};\
a\in {\cal E} \Leftrightarrow a=\sum_\tau \omega^\tau
\alpha_\tau,\ \omega^\tau \in \Omega (M),\ \alpha_\tau
\in {\cal A}.$

For gravitational theories it turns out that 
the local BRST cohomology in form degree $n$ and ghost number
$g$ can easily be obtained from the cohomology $H(s)$ of the
BRST differential $s$ \cite{BBH3}. Indeed, one has the following
isomorphism\footnote{This isomorphism holds because we assume
the spacetime manifold to be homeomorphic to ${\bf R}^n$.}:
\bb
H^n_g (s\vert d,{\cal E}) \simeq \frac{H^{g+n}(s,{\cal A})}{{\bf
R}}
\ee
In particular, this isomorphism holds in ghost number $0$ and
therefore for the consistent deformations. It is implemented
as follows: if $\alpha^n$ is a representative of $H^n(s,{\cal
A})$, then the corresponding representative of $H^n_0(s\vert
d,{\cal E})$ is  given by $a^n_0=\frac{1}{n!}b^n \alpha^n$ where
$b$ is the operator defined by $b=dx^\mu
\frac{\partial}{\partial
\xi^\mu}$ \cite{BBH3}.

The first part of the analysis of $H(s)$ consists in finding new
generators of the algebra ${\cal A}$ which isolate a
contractible part of the algebra with respect to the
differential $s$. Taking into account the results of
\cite{BBH3,Brandt93} we define the new basis of generators of
${\cal A}$ as:
\begin{eqnarray}
\{{\cal T}^r\}&=&\{D_{a_1}\ldots D_{a_k} R_{ab}^{\ \ I},\ 
D_{a_1}\ldots D_{a_k} H_{bcd}^{A}:
k=0,1,\ldots\},
\\
\{ {\cal T}^*_{\overline r}\} &=& \{ D_{a_1}\ldots D_{a_k}
{\hat\Phi}^*_\Gamma : k=0,1,\ldots \},\\
\hat\xi^a &=& \xi^\mu e_\mu^{\ a},\ \hat C^I=C^I + \xi^\mu
\omega_\mu^{\ I},
\\
\hat B^A &=& B^A + \xi^\mu B^A_\mu + \frac12 \xi^\nu \xi^\mu
B^A_{\mu\nu},\\
\{ U_l \} &=&\{ \partial_{(\mu_1\ldots \mu_k}e_{\mu)}^{\ a},\
\partial_{(\mu_1\ldots \mu_k}\omega_{\mu)}^{\ I}, \
\partial_{(\mu_1\ldots
\mu_k} B^A_{\nu_1)}, \ \partial_{(\mu_1\ldots
\mu_k} B^A_{[\nu_1)\nu_2]}:\nonumber \\ &&\quad
k=0,1,\ldots\},\\
\{V_l\} &=& \{ \partial_{(\mu_1\ldots \mu_k}se_{\mu)}^{\ a},\
\partial_{(\mu_1\ldots \mu_k}s\omega_{\mu)}^{\ I}, \
\partial_{(\mu_1\ldots
\mu_k} sB^A_{\nu_1)}, \ \partial_{(\mu_1\ldots
\mu_k} sB^A_{[\nu_1)\nu_2]}:\nonumber \\ &&\quad k=0,1,\ldots\},
\end{eqnarray} where,
\begin{eqnarray}
H^A_{abc}&=&E_{a}^{\ \mu}E_{b}^{\
\nu}E_{c}^{\ \rho}H^A_{\mu\nu\rho}
\\  R_{ab}^{\ I}&=&  R_{ab}^{\ \ cd}=E_{a}^{\
\mu}E_{b}^{\
\nu} R_{\mu\nu}^{\ \ cd},\\
\omega_\mu^{\ I}  &=& \omega_\mu^{\ ab},
\label{spincon}\\
\{ \hat\Phi^*_\Gamma \} &=&\{ \hat e_{a}^{*b},
\hat B^{*ab}_A,\hat \xi^*_a,\hat  C^*_I,
\hat B^{*a}_A,  \hat B^{*}_A\},\\
\hat e_{a}^{*b} &=&e_\mu^{\ b} e^{*\mu}_a/e,\ \hat C^*_I= 
C^*_I /e,\\
\hat B^{*ab}_A &=& e_\mu^{\ a} e_\nu^{\ b} B^{*\mu\nu}_A /e, \
\hat B^{*a}_A = - e_\mu^{\ a}B^{*\mu}_A /e,\ \hat B^{*}_A =
B^{*}_A/e \\
\hat \xi^*_a &=& E^{\ \mu}_a (\xi^*_\mu - A_\mu^{\ I} C^*_I -
B^A_{\mu
\nu} B^{*\nu}_A - B^A_{\mu} B^{*}_A)/e
\end{eqnarray}
It is easily shown that any local function can be expressed in
terms of the new generators.

The first advantage of the new basis is that on all but the
variables $U_l$ and $V_l$, the action of $s$ takes the familiar
form:
\begin{equation}
s=\delta + \gamma,
\end{equation}
with,
\bqn
&&\delta {\cal T}^r=\delta {\hat \xi}^a=\delta {\hat C}^I=\delta
\hat B^A=0,\label{eqSd}\\ &&\delta {\hat e}^{*b}_a=-(R_a^{\
b}-\frac12\delta_a^b
R)-\frac12H^A_{acd}H_A^{bcd}-\frac{1}{12}\delta_a^b
H^A_{cde}H_A^{cde},\\ &&\delta {\hat C}^*_{ab}=-2{\hat
e}^*_{[ab]},
\delta {\hat \xi}^*_a=-D_b{\hat e}^{*b}_a+\frac{1}{p_A!}{\hat
B}^{*b_1 b_{2}}_A {\hat H}^A_{a b_1 b_{2}}\\&&
\delta \hat{B}^{*a_{1}a_{2}}_{A}=D_b
\hat{H}_{A}^{ba_{1} a_{2}},
 \delta \hat{B}^{*a_{1}}_{A}=D_b\hat{B}^{*ba_{1}}_{A},
 \delta \hat{B}^{*}_{A}=D_b\hat{B}^{*b}_{A},
\\ &&
\delta D_{a_1}\ldots D_{a_k}{\hat\Phi}^*_\Gamma=D_{a_1}\ldots
D_{a_k}\delta{\hat\Phi}^*_\Gamma,
\eqn
\bqn
\gamma {\cal T}^r=({\hat \xi}^aD_a+{\hat C}^I\delta_I){\cal
T},\quad 
\gamma {\cal T}^*_{\overline r}=({\hat \xi}^aD_a+{\hat
C}^I\delta_I){\cal T}^*_{\overline r}\\
\gamma {\hat \xi}^a={\hat C}_b^{\ a}{\hat \xi}^b, \quad \gamma
{\hat C}^I=\frac{1}{2}{\hat C}^J{\hat C}^Kf_{KJ}^{\ \ I}+{\hat
F}^I, \quad \gamma {\hat B}^A= {\hat H}^A,\label{eqSf}
\eqn
where
\bb
{\hat R}^I=\frac{1}{2}{\hat \xi}^c {\hat \xi}^d R_{cd}^{\ \
I}, \quad {\hat H}^A=\frac{1}{6}{\hat
\xi}^{c}{\hat \xi}^{b} {\hat \xi}^{a} H^A_{abc}.
\ee 
In the above equations, the notation ${\hat C}^I\delta_I {\cal
T}^r$ stands for the Lorentz infinitesimal transformation of
${\cal T}^r$ with the infinitesimal parameter replaced by the
ghost ${\hat C}^I$. For example, if ${\cal T}^r={\cal T}^a$ is a
contravariant vector then ${\hat C}^I\delta_I {\cal
T}^a= {\hat C}^{a}_{\ b} {\cal T}^b$.

The gradings associated to $\delta$ and $\gamma$ are
respectively the {\em antighost} number (denoted {\em
antigh}) and the {\em pureghost} number (denoted {\em puregh});
their sum is equal to the {\em ghost} number (denoted {\em gh}).
The table summarizes the various gradings
associated to the operators, the fields, the ghosts and the
antifields.

\begin{table}[tbp]
\centerline{
\begin{tabular}{|c||c|c|c|c|}
\hline
 & parity & antigh & puregh & gh\\
\hline\hline
$s$ & 1 & - & - & 1 \\ \hline
$\delta$ & 1 & -1 & 0 & 1 \\ \hline
$\gamma$ & 1 & 0 & 1 & 1 \\ \hline\hline
${\cal T}^r$ & 0 & 0 & 0 & 0\\ \hline
${\hat \xi}^a$ & 1 & 0 & 1 & 1\\ \hline
${\hat C}^I$ & 1 & 0 & 1 & 1\\ \hline
${\hat B}^A$ & 0 & 0 & 2 & 2 \\ \hline
${\cal T}^*_{\overline r}$ & 1 & 1 & 0 & -1 \\ \hline
${\hat e}^{*b}_a$ & 1 & 1 & 0 & -1 \\ \hline
${\hat B}^{*ab}_A$ & 1 & 1 & 0 & -1 \\ \hline
${\hat \xi}^*_a$ & 0 & 2 & 0 & -2 \\ \hline
${\hat C}^*_I$ & 0 & 2 & 0 & -2 \\ \hline
${\hat B}^{*a}_A$ & 0 & 2 & 0 & -2 \\ \hline
${\hat B}^*_A$ & 1 & 3 & 0 & -3 \\ \hline
\end{tabular}
}
\label{variables}
\end{table}

The second advantage of the new basis is that it exhibits a
manifestly contractible part of the algebra ${\cal A}$.
Indeed, by construction, the variables $U_l$ and $V_l$ are
mapped on each other by the BRST differential,
\begin{equation}
sU_l=V_{l},\quad sV_{l}=0,
\end{equation}
and since the BRST differential does not mix the
$U_l$ and $V_l$ with the rest of the variables, each
couple $\{U_l,V_l\}$ drops out from the cohomology and we have
$H(s,{\cal A})\simeq H(s,{\cal A}_2)$ where ${\cal A}_2$ is the
algebra generated by the set $\{{\cal T}^r,{\cal T}^*_{\overline r},\hat\xi^a,\hat C^I,\hat
B^A\}$.

To summarize, the above discussion indicates that to obtain
$H^n_0(s\vert d,{\cal E})$ we only need to calculate
$H^n(s,{\cal A}_2)$. This is the subject of the next section.

\section{BRST cohomology in ${\cal{A}}_2$ and consistent
vertices}
\setcounter{equation}{0}
\setcounter{theorem}{0}

In order to get the cohomology $H^n(s,{\cal A}_2)$ we need
to solve the equation,
\bb 
s\alpha^n=0, \quad \alpha^n \in {\cal A}_2,\label{eqS}
\ee
where two solutions of (\ref{eqS}) are identified if they
differ by an $s$-exact contribution, i.e, $\alpha^n
\sim \alpha^n + s\beta^{n-1}$ with $\beta^{n-1}\in {\cal A}_2$.

The approach we follow is identical to the one developed in
\cite{BBH3} so we only emphasize here the main ideas and the new
results. A more in-depth presentation will be given in
\cite{BKS2}.

First, the cocycle $\alpha^n$ is decomposed according to a
degree called the
$\hat\xi^a$-degree which counts the polynomial degree in the
variables $\hat\xi^a$,
\bb 
\alpha^n = \sum_{k=l}^n \alpha_k.
\ee
According to the $\hat\xi^a$-degree, the BRST differential
decomposes into four parts, $s=s_0+s_1+s_2+s_3$, which can be
read off from (\ref{eqSd})-(\ref{eqSf}). The first term is given
by,
\bb
s_0=\delta + \gamma_L
\ee
where $\delta$ is the Koszul-Tate differential and $\gamma_L$
is the longitudinal exterior derivative along the gauge
orbits of ${\cal G}_L$,
\bb
\gamma_L=-\frac12 \hat C^J \hat C^K f_{JK}^{\ \
I}\frac{\partial}{\partial \hat C^I}+\hat C^I \delta_I,
\ee
with,
\bb 
\delta_{ab}\hat\xi_c=\eta_{bc}\hat\xi_a-\eta_{ac}\hat
\xi_b,\quad \delta_I \hat C^J=-f_{IK}^{\ \ J}\hat C^K.
\ee
$s_1$ plays the r\^ole of an exterior covariant derivative whose
differentials are the $\hat\xi^a$. Its action is given by,
\bb
s_1 {\cal T}^r =\hat\xi^a D_a {\cal T}^r, \ s_1 {\cal
T}^*_{\overline r}= \hat\xi^a D_a {\cal T}^*_{\overline r}, \ 
s_1 \hat\xi^a=s_1 \hat C^I=s_1\hat B^A=0.
\ee
Finally, the operators $s_2$ and $s_3$ are given by,
\bb
s_2 = \hat R^I \frac{\partial}{\partial \hat C^I},\quad
s_3 = \hat H^A \frac{\partial}{\partial \hat B^A}.
\ee
According to the $\hat\xi^a$-degree, eq. (\ref{eqS}) decomposes
into the following tower of equations:
\bqn
0&=&s_0 \alpha_l,\label{des1}\\
0&=&s_0 \alpha_{l+1}+s_1\alpha_l,\label{des2}\\
0&=&s_0 \alpha_{l+2}+s_1\alpha_{l+1}+s_2\alpha_l,\\
&\vdots&\nonumber
\eqn
Up to trivial terms which only modify components of higher
$\hat\xi^a$-degree, eq. (\ref{des1}) indicates that $\alpha_l$
is an element of the cohomology $H(s_0,{\cal A}_2)$.
Given the definition of $s_0$, this cohomology is analyzed in a
very similar fashion as the standard BRST cohomology for
non-gravitational theories. Using the acyclicity of $\delta$ in
antighost number $k>0$, one can show that each class of
$H(s_0,{\cal A}_2)$ admits an antifield independent
representative so that the cocycle condition becomes, $\gamma_L
\alpha_l=0$. This equation is the well known coboundary
condition for the Lie algebra cohomology of
${\cal G}_L$ in a ${\cal G}_L$-module so the most general form
for the non-trivial
$\alpha_l$ is, 
\bb
\alpha_l=\alpha^i_l(\hat\xi^a,{\cal T}^r)\omega_i(\theta_K(\hat
C^I), \hat B^A), \quad \delta_I \alpha^i_l(\hat\xi^a,{\cal
T}^r)=0.\label{alphal}
\ee
In (\ref{alphal}), the $\omega_i$ are polynomials in the $\hat
B^A$ and the
$\theta_K(\hat C^I)$ which are the primitive elements of the
Lorentz Lie algebra cohomology.
In
$n=2r$ and $n=2r+1$ dimensions, they are given by,
\begin{eqnarray}
\theta_{K}(C)&=&C_{a_1}^{\ a_2}D_{a_2}^{\ a_3}\ldots 
D_{a_{2K}}^{\ \ \ a_1},\quad
K=1,2,\ldots,r-1\ , \label{th1}\\
\theta_r(C)&=&\left\{\begin{array}{ll}
C_{a_1}^{\ a_2}D_{a_2}^{\ a_3}\ldots D_{a_{2r}}^{\ \ a_1} &
\mbox{for}\ n=2r+1 \\
\epsilon_{a_1 b_1\ldots a_r b_r}C^{a_1 b_1}
D^{a_2 b_2}\ldots D^{a_r b_r} & \mbox{for}\ n=2r \label{th2}
\end{array}\right.
\end{eqnarray}
where $D_a{}^b = C_a{}^cC_c{}^b$. 

Let us first consider the case $l=n$. Since we are interested
in solutions of (\ref{eqS}) in ghost number $n$, in
(\ref{alphal}) we necessarily have $\omega_i (\theta_K(\hat
C^I), \hat B^A)=k_i$ where the $k_i$ are constants (the
$\hat\xi^a$ are of ghost number 1). In that case, no further
condition is imposed on $\alpha_n$ and we have,
\bb 
\alpha=\alpha_n= L({\cal T}) \hat\Theta, \quad \delta_I L=0,
\ee
where $\hat \Theta= \hat \xi^0 \ldots \hat\xi^{n-1}$. The above
elements of $H(s)$ give rise in $H^n_0(s\vert d,{\cal E})$ to
the cocycles,
\bb
\alpha= L({\cal T}) d^nx, \quad \delta_I L=0,
\ee
which constitute the first set of consistent interactions
announced in the introduction. Since these vertices are
strictly gauge invariant they do not require any modification in
the gauge transformations of the theory. 

We now turn our attention to the case $l<n$. To that end, we
substitute the general form (\ref{alphal}) into eq.
(\ref{des2}). By doing so, one easily proves \cite{BBH3} that
$\alpha^i_l(\hat\xi^a,{\cal T}^r)$ has to obey the following
equation,
\bb 
s_1 \alpha_l^i+\delta \alpha^i_{l+1,1}=0,\label{eqchar}
\ee
where $\alpha^i_{l+1,1}=\alpha^i_{l+1,1}(\hat\xi^a,{\cal
T}^r,{\cal T}^*_{\overline r})$ is of antighost number $1$ and
satisfies $\delta_I \alpha^i_{l+1,1}(\hat\xi^a,{\cal
T}^r,{\cal T}^*_{\overline r})=0$. Trivial
solutions of (\ref{eqchar}) ($\alpha^i_l= s_1 \beta^i_{l-1}+
\delta \beta^i_l$) are irrelevant since they amount to trivial
contributions in $\alpha$.

Eq. (\ref{eqchar}) along with its coboundary condition define
the so-called ``invariant characteristic cohomology"
$H^{inv}_{char}(d)$ (with $d$ formally substituted by $s_1$)
which plays a central r\^ole in the analysis of any local field
theory. Theorems concerning
$H^{inv}_{char}(d)$ in the cases of pure gravity and $p$-form
gauge theory can be found respectively in \cite{Barnich:1995db}
and
\cite{HKS1}. The extension of those results in the
case of gravity coupled to a system of $2$-forms is fully
treated in \cite{BKS2}. 
Here, we may restrict our attention to solutions of
eq. (\ref{eqchar}) in ghost number $\leq n-2$. Indeed,
taking into account (\ref{th1}) and (\ref{th2}) we see that the
$\theta_K(\hat C^I)$ and $\hat B^A$ are at least of ghost number
$2$; therefore, to construct solutions of (\ref{eqS}) in ghost
number $n$, we necessarily have $l\leq n-2$. In that case, the
most general solution of (\ref{eqchar}) is up to trivial
terms given by \cite{BKS2},
\bb
\alpha_l^i=P^i(f_K,\hat H^A)+ \delta_l^{n-3} k^i_A {\overline
{\hat H}}^A+ \delta_l^{n-2}\delta^n_4 k^i_{AB} {\overline {\hat
H}}^A {\overline {\hat H}}^B.
\label{solchar}
\ee
In (\ref{solchar}), the $k^i_A$ and $k^i_{AB}=- k^i_{BA}$ are
constants, ${\overline {\hat H}}^A$ is the Hodge dual of $\hat
H^A$ and the $f_K$ are generators for the the
${\cal G}_L$-invariant polynomials in the $\hat R^I$. The
$f_K$
are given in the cases
$n=2r$ and $n=2r+1$ by:
\begin{eqnarray}
f_K&=&
{\hat R}_{a_1}^{\ a_2}{\hat R}_{a_2}^{\ a_3}\ldots
{\hat R}_{a_{2K}}^{\ \ a_1},\quad K=1,2,\ldots,r-1\ ,\\
f_{r}&=&\left\{\begin{array}{ll}
{\hat R}_{a_1}^{\ a_2}{\hat R}_{a_2}^{\ a_3}\ldots
{\hat R}_{a_{2r}}^{\ \ a_1}
& \mbox{for}\ n=2r+1 \\ \epsilon_{a_1b_1\ldots a_rb_r}
{\hat R}^{a_1b_1}\ldots{\hat R}^{a_rb_r}
& \mbox{for}\ n=2r.
\end{array}\right.
\end{eqnarray}
{}From (\ref{solchar}) we see that $\alpha^i_l$ involves
two kinds of contributions. The first consists of the
polynomials $P^i(f_K,\hat H^a)$ which obey (\ref{eqchar})
without the need for a term $\alpha^i_{l+1,1}$. One says that
the corresponding $\alpha^i_l$ are {\em strongly} $s_1$-closed.
On the other hand, the last two terms of (\ref{solchar}) are
only {\em weakly}
$s_1$-closed since they require a term $\alpha^i_{l+1,1}$ in
order to satisfy eq. (\ref{eqchar}).

Let us first consider the BRST cocycles generated by the
strongly $s_1$-closed $\alpha^i_l$. Their part of lowest
$\hat\xi^a$-degree is of the form,
\bb
\alpha_l = P^i(f_K,\hat H^A)\omega_i(\theta_K(\hat
C^I), \hat B^A).\label{RRR}
\ee
We need to complete these $\alpha_l$ by terms of higher $\hat
\xi^a$-degree in order to obtain BRST cocycles. To do this, we
associate to the $\theta_K$ the following quantities
\cite{BBH3}:
\bb
q_K=\sum_{k=0}^{2K-1}(-)^k\frac{(2K)!(2K-1)!}{(2K+k)!(2K-k-1)!}\,
Str({\hat{\cal C}}{\hat{\cal D}}^k{\hat{\cal
R}}^{2K-k-1}),\label{stalpha}
\ee
with,
\bb
\hat{\cal C}=\hat C^I T_I,\quad \hat {\cal D}=\hat {\cal C}^2,
\quad \hat{\cal R}=\hat R^I T_I.\label{repr}
\ee
In (\ref{repr}), the $T_I$ are the matrices of the adjoint
representation of $so(n-1,1)$ except for $n=2r$ in which the
spinor representation is used. At lowest $\hat\xi^a$-degree the
$q_K$ begin with $\theta_K$ and are such that $sq_K=f_K$.

The usefulness of the $q_K$ stems from the fact that if
$\alpha$ is a cocycle with $\alpha_l$ as in
(\ref{RRR}), then one may assume that
$\alpha=\alpha(f_K,q_K,\hat H^A,\hat B^A)= P^i(f_K,
\hat H^A)\omega_i(q_K, \hat B^A)$, i.e., $\alpha$ is obtained
from
$\alpha_l$ by replacing the $\theta_K$ by the $q_K$
\cite{BKS2}. Furthermore, one can show 
that if $\alpha=\alpha(f_K,q_K,\hat H^A,\hat B^A)=s\beta$
then $\beta=\beta(f_K,q_K,\hat H^A,\hat B^A)$. 
The analysis of the BRST cocycles arising from (\ref{RRR})
is therefore reduced to the investigation of the BRST cohomology
in the so-called small
algebra generated by the variable $q_K,f_K,\hat
H^A,\hat B^A$. 

This problem has been studied in detail for the pure
gravitational case in \cite{Brandt:1990et}. It is shown that
in ghost number $n$ the BRST cocycles in the small algebra
are of the ``Chern-Simons" form, $\alpha=q_K P(f_K)$.
Here,
the only difference comes from the presence of $\hat H^A$ and
$\hat B^A$ among the generators of the small algebra; they are
related by
$s\hat B^A=\hat H^A$. The procedure is nearly identical to
\cite{Brandt:1990et} and one can show
\cite{BKS2} that in ghost number
$n$ the BRST cocycles are again of the Chern-Simons form, i.e.,
\bb
\alpha=q_K P_K(f_L,\hat H^A)+\hat B^A Q_A(f_K,\hat
H^B).\label{CCC}
\ee
The corresponding consistent deformations
are obtained from (\ref{CCC}) by the substitution
$\hat\xi^a\rightarrow dx^\mu$. They constitute the second set
of vertices announced in the introduction. Note that they are
invariant up to a boundary term and therefore do not require a
modification of the gauge transformations.

We now consider the BRST cocycles generated by the weakly
$s_1$-closed $\alpha^i_l$. Taking into account
(\ref{th1}), (\ref{th2}) and (\ref{solchar}), we see that in
spacetime dimension 
$n>4$, the only possibility we have in ghost number $n$ for
$\alpha_l$ is,
\bb
\alpha_{n-3}=k^1_{A}{\overline {\hat H}}^A \theta_1,\label{NN4}
\ee
with $\theta_1=C_a^{\ b} C_b^{\ c} C_c^{\ a}$. $\alpha_{n-3}$
has to be completed with terms of higher $\hat
\xi^a$-degree in order to obtain a BRST cocycle. 
To that end we introduce the following notation \cite{HKS1}:
\bb
q^{*A}= {\overline {\hat H}}^A+ {\overline {\hat B}}^{*A}_1
+ {\overline {\hat B}}^{*A}_2 + {\overline {\hat B}}^{*A}_3,
\ee
where the ${\overline {\hat B}}^{*A}_j$ are defined through,
$\delta {\overline {\hat B}}^{*A}_1 + d{\overline {\hat
H}}^{A}=0$, $\delta {\overline {\hat B}}^{*A}_2 + d{\overline
{\hat B}}^{*A}_1=0$, $\delta {\overline {\hat B}}^{*A}_3 +
d{\overline {\hat B}}^{*A}_2=0$ (${\overline {\hat
B}}^{*A}_j$ is proportional to the dual of the antifield of
antighost $j$, with the $\hat\xi^a$ as differentials).

Using the above definition, it is straightforward to complete 
(\ref{NN4}) to a BRST cocycle. Indeed, as can be seen by direct
substitution,
\bqn
\alpha&=&k^1_A q^{*A}q_1\label{ver1}\\
&=&k^1_{A}{\overline {\hat H}}^A
\theta_1+\alpha_{n-2}+\alpha_{n-1}+\alpha_n.
\eqn
is a solution of (\ref{NN4}). The corresponding
elements of $H(s\vert d,{\cal E})$ are again obtained from
(\ref{ver1}) by the substitution
$\hat\xi^a\rightarrow dx^\mu$.
Their components of antighost number zero give rise to the
following class of consistent vertices:
\bb
V=k^1_A {\overline H}^A Tr(\omega
R-\frac{1}{3}\omega^3)\label{vertex}
\ee
where ${\overline H}^A$ is the dual of the field strength $H^A$,
$\omega=\omega^I_\mu dx^\mu T_I$ and $R=\frac12dx^\mu dx^\nu
R_{\mu\nu}^{\ \ I} T_I$ (the $T_I$ are the matrices of the
adjoint representation of $so(n-1,1)$).  These vertices require
a modification in the gauge transformations since they are not
invariant under the original ones. They belong to the third type
of interactions described in the introduction.

In the particular case of spacetime dimension $n=4$ 
further couplings are possible. Indeed, in ghost number $n=4$ we
have the following candidates for $\alpha_l$,
\bb
\alpha_{l}=k^1_{A}{\overline {\hat H}}^A \theta_1 +
k^2_{A}{\overline {\hat H}}^A \theta_2 +k_{ABC}{\overline {\hat
H}}^A {\overline {\hat H}}^B \hat B^C,
\ee
where $k_{ABC}=-k_{BAC}$ and
$\theta_2=\epsilon_{abcd}C^{ab}C^{ce}C_{e}^{\ d}$. The above
$\alpha_l$ are easily completed to the following
BRST-cocycles,
\bb 
\alpha=k^1_A q^{*A}q_1 +k^2_A q^{*A}q_2+
k_{ABC}q^{*A}q^{*B}\hat B^A.\label{mkl}
\ee
The first two terms yield consistent vertices identical to
(\ref{vertex}) (with the trace taken once in the adjoint
representation and once in the spinorial representation of
$so(n-1,1)$):
\bb
V=k^1_A {\overline H}^A Tr_{adj}(\omega
R-\frac{1}{3}\omega^3)+k^2_A {\overline H}^A Tr_{sp}(\omega
R-\frac{1}{3}\omega^3)
\ee
The last term in (\ref{mkl}) produces the Freedman-Townsend
coupling,
\bb
V=k_{ABC}{\overline H}^A {\overline H}^B B^C.
\ee
This vertex is again not invariant under the original gauge
transformations and a modification of these transformations is
imposed.

\section{Comments}
In this article we have computed the local BRST cohomology
$H(s\vert d)$ in ghost number $0$ and form degree $n$ in order
to obtain all the first-order vertices of gravity coupled to a
system of
$2$-form gauge fields. 

The first two types of couplings we obtain (strictly invariant
polynomials in the curvatures and Chern-Simons forms) are
consistent to all orders in the coupling constant since they
are gauge invariant (up to a total derivative for the
Chern-Simons forms) under the original gauge transformations.
These vertices correspond to antifield independent
representative of the BRST cohomology.

The last two types of vertices we describe are not
invariant under the original gauge transformations since
they depend non-trivially on the antifields. In that case a
modification of the gauge transformations is required as well
as the addition of higher-order vertices in order to obtain a
theory consistent to all orders in the coupling constant. For
the exotic couplings of the 2-form gauge fields to
gravitational Chern-Simons forms, the full theory is of the
Chapline-Manton type in which the 2-form curvature present in
(\ref{lagrangian}) is replaced by $H'_A=H_A+
k_A Tr(\omega F-\frac13\omega^3)$. The Freedman-Townsend
coupling gives rise to a non-polynomial full theory.
Polynomiality can be restored by describing the theory with a
``first-order formulation" and the full lagrangian is simply a
covariantized version of the original Freedman-Townsend theory.

Our analysis can be extended to cover the couplings
of gravity to 
$p$-forms \cite{BKS2}.

\section*{Acknowledgements}
We are grateful to  Marc Henneaux for suggesting the
problem. We also thank him along with Glenn Barnich and
Friedemann Brandt for useful discussions. This work is supported
in part by the ``Actions de Recherche Concert{\'e}es" of the
``Direction de la Recherche Scientifique - Communaut{\'e}
Fran{\c c}aise de Belgique" and by IISN - Belgium (convention
4.4505.86).

\end{document}